\documentstyle[aps, twocolumn]{revtex}  
\begin{document}
\twocolumn[\begin{center}
{\huge \bf Spectrum of the 2-Flavor Schwinger Model
from the Heisenberg Spin Chain}\\
\vskip 0.2 truein
{\bf F. Berruto$^{(a)}$, G. Grignani$^{(a)}$, G. W. Semenoff$^{(b)}$
and P.  Sodano$^{(a)}$}
\vskip 0.1truein
$(a)$ {\it Dipartimento di Fisica and Sezione
I.N.F.N., Universit\`a di Perugia, Via A. Pascoli I-06123 Perugia,
Italy}\\
\vskip 0.1truecm
$(b)$ {\it Department of Physics and Astronomy, University of British
Columbia, 6224 Agricultural Road,\\Vancouver, British Columbia, Canada
V6T 1Z1}\\
\vskip 0.1truein
\end{center}
We study the strong coupling limit of the 2-flavor lattice Schwinger
model in the Hamiltonian formalism using staggered fermions.  We show
that the problem of finding the low lying states is equivalent to
solving the Heisenberg antiferromagnetic spin chain.  We find good
agreement with the continuum theory.\\
\vskip 0.1truein] Very appealing is the idea that solvable models for
simple spin-$\frac{1}{2}$ quantum antiferromagnets are related to
quantized gauge theories.  In particular, we think it is interesting
to exhibit a gauge theory with the same low-lying excitations of the
antiferromagnetic Heisenberg chain, since this system admits
spin-$\frac{1}{2}$ quark-like spinon excitations, but has physical
states with integer spin and an even number of spinons~\cite{faddeev}.
The main purpose of this Letter is to establish the correspondence
between the low-lying excitations of the antiferromagnetic Heisenberg
chain and those of the 2-flavor Schwinger model.

A relation between quantum
antiferromagnets and gauge theories is widely known to appear in
various contexts.  For example, a
mechanism of confinement for spin-Peierls antiferromagnets was
introduced in \cite{aff}.  A related idea has been pursued by
Laughlin~\cite{BL} who argued that analogies exist between spectral
data of gauge theories and of strongly correlated electron systems.
The analogy has also been used to discuss spin ladders~\cite{hos}.

In the context of conventional lattice gauge theory, it is known that
features of the strong coupling limit, particularly those involving
chiral symmetry breaking, have analogs in quantum spin
systems~\cite{smit}.  In many cases, the problem of finding the ground
state of the strongly coupled lattice gauge theory is equivalent to
finding the ground state of a generalized quantum
antiferromagnet~\cite{sem}.  This could
possibly be exploited to obtain non-perturbative information about the
structure of the gauge theory states.

This approach suffers from the fact that strong coupling limits of
gauge theories are highly non-universal.  In four or less dimensions,
the continuum limit and the universal behavior occur at weak coupling.
There are many choices of strong coupling theory which should produce
identical continuum physics.  In spite of this difficulty, there exist
strong coupling computations which claim some degree of
success~\cite{kog}.  In a previous paper~\cite{noi} we re-examined
the lattice Schwinger model using staggered fermions in the
Hamiltonian formulation. The staggered fermions possess a discrete
chiral symmetry and we showed how chiral symmetry breaking, which
arises from the axial anomaly in the continuum theory, is the result
of spontaneous symmetry breaking in the strong coupling limit.  The
solution of the strong coupling problem was identical to solving a
particular type of Ising system with long-ranged interaction. We
showed that the mass of the elementary
excitation and the chiral condensate could be computed reliably from
an extrapolation to weak coupling using Pad\'e approximants. Our
analysis improved previous ones by taking careful account of all
discrete symmetries of the continuum theory.

In this Letter, we shall test these ideas further by examining the
ground state and low energy excitations of the strongly coupled
2-flavor Schwinger model.  We make the point that, even with its
inherent nonuniversality, the most straightforward strong coupling
limit makes accurate predictions of the masses and quantum numbers of
the known elementary excitations of the continuum model.  It also
gives an intuitive picture of the nonperturbative features of the
ground state.

We show that to find the strong coupling ground state, we must solve
for the ground state of the Heisenberg antiferromagnetic spin chain.
This state and its excitations are known explicitly from the Bethe
ansatz \cite{bethe}.  We show that these states have the spectrum and
quantum numbers which are expected for the ground state and massless
excitations of the 2-flavor Schwinger model in the continuum.  We
stress that this is a non-trivial correspondence which would not be
easy to find at weak coupling.  Chiral symmetry breaking in the
Schwinger model would be identical to formation in the antiferromagnet
of either commensurate density waves when
$\left<\bar\psi\psi\right>\neq 0$ or antiferromagnetic order when
$\left< \bar\psi \sigma^a\psi\right>\neq 0$.  The Bethe ansatz
solution of the antiferromagnet shares the property of the 2-flavor
Schwinger model that there are no condensates of this form. The lowest
lying excitations of the antiferromagnet are a singlet and a
triplet~\cite{faddeev}. We shall show that they have identical parity
and G-parity to the lowest lying excitations of the continuum 2-flavor
Schwinger model ~\cite{coleman}.

Aside from the massless excitations, we find good agreement at strong
coupling with the quantum numbers and mass of the
Schwinger model boson, which we find is created from the Bethe ansatz ground
state by charge transport.  We show that essential quantities in the
strong coupling expansion can be expressed in terms of spin-spin
correlators of the quantum antiferromagnet.

The Hamiltonian, gauge constraint and non-vanishing (anti-)commutators
of the continuum 2-flavor Schwinger model are
\begin{eqnarray}
H=\int dx[\frac{e^2}{2}E^2(x)+\sum_{a=1}^2&
\psi^{\dagger}_a (x)\alpha\left(i\partial_x +eA(x)\right)\psi_a
(x)]\nonumber\\ \partial_x E(x)\ +&\sum_{a=1}^2 \psi^{\dagger}_a
(x)\psi_a (x)\sim 0\nonumber\\
\left[ A(x),E(y)\right]=i\delta(x-y),&
 \left\{\psi_a(x),\psi_b^{\dagger}(y)\right\}=\delta_{ab}\delta(x-y)\nonumber
\end{eqnarray}
A lattice Hamiltonian, constraint and (anti-) commutators that reduce
to these in the naive continuum limit are 
\begin{eqnarray}
H_{S}=\frac{e^{2}a}{2}\sum_{x=1}^N E_{x}^{2}&-&\frac{it}{2a}\sum_{x=1}^N
\sum_{a=1}^2 (\psi_{a,x+1}
^{\dag}e^{iA_{x}}\psi_{a,x}-h.c.)\nonumber\\
E_{x}-E_{x-1}&+&\psi_{1,x}^{\dag}\psi_{1,x}+\psi_{2,x}^{\dag}\psi_{2,x}-1\sim
0\ ,
\label{gauss}\\
\left[ A_x,E_y\right]=i\delta_{x,y}~&,&
\left\{\psi_{a,x},\psi_{b,y}^{\dagger}\right\}=\delta_{ab}\delta_{xy}
\nonumber
\end{eqnarray}
where the fermion fields are defined on the sites, $x=1,...,N$, the
gauge and the electric fields, $A_{x}$ and $E_{x}$, on the links $[x;
x + 1]$, $N$ is an even integer and, when $N$ is finite, periodic
boundary conditions should be used.  When $N$ is finite, the continuum
limit is the two flavor Schwinger model on a circle.  In
Eq.(\ref{gauss}) the constant $1$ has been subtracted from the
naively defined charge density operator, in order to make the gauge
generator odd under the usual charge conjugation transformation. The
properly defined charge density reads
\begin{equation}
\rho(x)=\psi_{1,x}^{\dag}\psi_{1,x}+\psi_{2,x}^{\dag}\psi_{2,x}-1
\label{charged}
\end{equation}
and vanishes on every site occupied by only one
particle. 

The lattice 2-flavor Schwinger model is equivalent to a one
dimensional quantum Coulomb gas on a lattice with two different kinds
of particles. To see this we fix the Coulomb gauge, $A_{x} = A$.
Eliminating the non-constant electric field using the gauge
constraint, one obtains the effective Hamiltonian
where the constant 
modes of the gauge field decouple in the thermodynamic limit
$ N \longrightarrow \infty $ \cite{noi}.
In this limit
the Schwinger Hamiltonian, rescaled by the factor ${e^{2}a}/{2}$,
becomes $H=H_{0}+\epsilon H_{h}$, where
\begin{eqnarray}
H_{0}&=&\sum_{x>y}[\frac{(x-y)^{2}}{N}-(x-y)]\rho(x)\rho(y)\ ,
\label{hu}\\
H_{h}&=&-i(R-L)
\label{hp}
\end{eqnarray}
and $\epsilon=t/e^{2}a^{2}$.
In Eq.(\ref{hp}) the right $R$ and left $L$ hopping operators
have been defined ($L=R^{\dagger}$)
$$
R=\sum_{x=1}^{N}R_{x}=\sum_{x=1}^{N}\sum_{a=1}^{2}R_{x}^{(a)}= 
\sum_{x=1}^{N} \sum_{a=1}^2 \psi_{a,x+1}^{\dag}e^{iA}\psi_{a,x}
$$
On a periodic chain they commute, $[R,L]=0$.

We shall consider a strong coupling perturbative expansion where
$H_0$ is the unperturbed Hamiltonian and $H_p$ the
perturbation. The $H_0$ ground state is highly degenerate. Due
to (\ref{charged}), every state with one particle per site,
has zero energy.  There are $2^N$ states of this type.

First order perturbations to the vacuum energy vanish.
The leading term in the vacuum energy is  
of order $\epsilon^2$ 
\begin{equation}
E^{(2)}_{0}=<H_{h}^{\dagger}\frac{\Pi}{E_{0}^{(0)}-H_{0}}H_{h}>
\label{secorder}
\end{equation}
where the expectation values are defined on the degenerate subspace
and $\Pi$ is a projection operator which projects orthogonal to the set
of states with one particle per site.  Due to the vanishing of the
charge density on the ground states of the $H_{0}$, the commutator
$
[H_0, H_h]=H_h
$
holds on any linear combination of the degenerate ground states. 
Consequently, from Eq.(\ref{secorder}) one finds
\begin{equation}
E^{(2)}_{0}=-2<RL>
\label{secorder2}
\end{equation}
On the ground state the combination $R L$ can be reexpressed in terms of the 
Heisenberg Hamiltonian.
By introducing the Schwinger spin operators
\begin{equation}
\vec{S}_{x}=\psi_{a,x}^{\dag}\frac{\vec{\sigma}_{ab}}{2}\psi_{b,x}
\end{equation}
the Heisenberg Hamiltonian $H_{J}=\sum_{x=1}^{N}(\vec{S}_x
\cdot \vec{S}_{x+1}-\frac{1}{4})$ reads
\begin{equation}
H_{J}=-
\sum_{x=1}^{N} (\frac{1}{2} 
L_{x}R_{x}+\frac{1}{4}\rho(x) \rho(x+1) )
\end{equation}
so that  on the degenerate subspace we have 
\begin{equation}
<H_{J}>=<\sum_{x=1}^{N}(-\frac{1}{2}L_{x}R_{x})>
\label{mainequation}
\end{equation}
Taking into account that products of $L_x$ and $R_y$ at different points 
have vanishing expectation values on the ground states,
and using Eq.(\ref{mainequation}), Eq.(\ref{secorder2}) reads
\begin{equation}
E^{(2)}_{0}=4<H_{J}>
\label{secorder3}
\end{equation}
The problem of determining the correct ground state, on which to 
perform the perturbative expansion, 
is then reduced to the diagonalization of the Heisenberg spin 
Hamiltonian. As is well known, in one dimension $H_{J}$ 
is exactly diagonalizable 
\cite{bethe,faddeev}. 
 
On a given site, the presence of a flavor $1$ particle 
can be represented, in the spin model, by the presence of a spin up, 
a flavor $2$ particle by the presence of a spin down. 
The number of the $H_J$ eigenstates is $2^N$.
Among these, the spin singlet with lowest energy is the non degenerate
ground state $|g.s.>$.
Consequently, $|g.s.>$ is the ground state that must be used 
in the strong coupling perturbation theory of the two flavor Schwinger model.
$|g.s.>$ is translationally invariant, namely it is invariant 
under the discrete chiral symmetry \cite{noi}. Therefore, at variance
with the one flavor case, the chiral symmetry cannot 
be spontaneously broken even in the infinite coupling limit.
$|g.s.>$ has vanishing charge density 
on each site, so that it does not support any electric flux 
$\rho(x)|g.s.>=0\ ,\ E_{x}|g.s.>=0 \ (x=1,...,N)$.
$|g.s.>$ is expressed as a linear combination of all the states with
$N/2$ spins up and $N/2 $ spins down. The eigenvalue of the Heisenberg
Hamiltonian on this state in the thermodynamic limit is known
~\cite{bethe,faddeev},
$
H_{J}|g.s.>=(-N\  \ln\ 2)|g.s.>
$,
and it provides
the second order correction Eq.(\ref{secorder3}), 
$E_{g.s.}^{(2)}=-4 N\ln 2$, to the vacuum energy.

We now show that it is possible to identify the low lying excitations
of the Schwinger model with those of the Heisenberg model and that the
mass gaps of any other excitation can be expressed as functions of
v.e.v.'s of powers of $H_{J}$ and spin-spin correlation functions.

The continuum 2-flavor Schwinger model excitations are characterized
by the quantum numbers of $P$- and $G$-parity. For the massive
model, in the limit where the mass of the fermions is small compared
to $e^2$ (strong coupling), Coleman showed~\cite{coleman} that the
lightest state is $I^{PG}=1^{-+}$ and the next state up is
$I^{PG}=0^{++}$.  The Schwinger model in this limit is equivalent to a
sine-Gordon in $d=1+1$ and the low lying states can be
interpreted as soliton-antisoliton states. There are also other states
in the theory, way up in mass, $e.g.$ a pseudoscalar isosinglet
$I^{PG}=0^{--}$, G-odd with mass $m=\sqrt{2/ \pi} \ e$.  All these
excitations can be reobtained on the lattice. In the limit of
vanishing fermion mass the first two correspond to massless Heisenberg
states, the massive states are instead generated by fermion
transport.  

As we have shown, the Schwinger model vacuum, is the ground state of
the antiferromagnet, $|g.s.>$, thus any excitation should be
created from this state.  There are two different types of excitations
that can be created from $|g.s.>$. Those that involve only spin
flipping and those that involve fermion transport besides spin
flipping.  The excitations of the first type have lower energy since
no electric flux is created, those of the second type have a higher
energy and the lowest energy ones occur when the fermion is
transported a minimal distance, since the energy is proportional to
the coupling times the length of the electric flux that is created.
Only the first type of excitations can be described in terms of the
Heisenberg model excited states. In \cite{faddeev} a complete
classification of the Heisenberg model excitations has been given.  
It was shown
that any excitation can be interpreted as the scattering of
quasiparticles of spin 1/2, and that in physical states there is only
an even number of these kink-spin waves, so that there are only states
with integer spin. In the thermodynamic limit the two lowest
excitations are a triplet and a singlet, they have a dispersion
relation depending on 2 parameters, (the momenta of the two kinks) and
for vanishing total momentum (relative to the ground state momentum
$P_{g.s.}=0$ for $N/2$ even, $P_{g.s.}=\pi$ for $N/2$ odd) they are
degenerate with the ground state.  This interpretation is consistent
with the one given by Coleman in terms of solitons in the Schwinger
model.  In fact these excitations also have the correct Schwinger
model quantum numbers (the spin becomes the isospin) and energies, so
that they can be identified with the lowest lying excitation for this
model in strong coupling.  There is a whole set of excitation of this
type (in the notation of Ref.\cite{faddeev}, all those belonging to
the class $\cal{M}_{AF}$) all of them, for zero momentum in the
thermodynamic limit, would be gapless.  They are eigenstate of the
total momentum. A G-parity transformation corresponds to a translation
by one site, therefore at zero momentum they have a positive G-parity
(with respect to the G-parity of the ground state) as expected for the
low lying Schwinger model states.  For finite systems these
excitations are gapped so that the P-parity is a well defined quantum
number. The P-parity and spin of the lowest lying excitations has been
considered in
\cite{eggert} and the two lowest-lying states were shown to be $s^{P}=1^{-}$
and $s^{P}=0^{+}$, $i.e.$ with the parity of the lowest-lying states
found by Coleman in the Schwinger model.

Let us discuss their energies.  The excitation masses are given by the
difference between the energy of the excitation and the energy of the
ground state at zero momentum.  The strong coupling perturbative
expansion for the energies of any of the state belonging to
$\cal{M}_{AF}$, which we denote by the generic symbol $|ex>$, have the
same expression as those of the ground state.  Consequently if the
states $|ex>$ are taken at zero momentum, up to the second order in
the strong coupling expansion, they have the same energy of the ground
state (\ref{secorder}), $E^{(2)}_{ex}=-4 N\ln 2$.  To this order the
mass gap is zero. At higher orders a mass gap might arise, due to the
fact that correlators of spin products $\vec{S}_x\cdot\vec{S}_{x+d}$
with $d>1$ arise and these might be different on the ground state and
on the excited states. The values of these correlators is not known,
only on the ground state they have been considered in
\cite{korepin} and their asymptotic behavior has been computed in
\cite{coraff}. 
The lowest lying among the state $|ex>$, have the correct quantum numbers,
are gapless, at least up to second order in the strong coupling expansion, 
and can then be identified with the lowest lying excitations of the
Schwinger model.

Let us now consider the states obtained by fermion transport
of one site on the Heisenberg model ground state. Using the spatial
component of the Schwinger model currents two states can be created,
a pseudoscalar isosinglet $I^{PG}=0^{--}$, G-odd, and a
pseudoscalar isotriplet $I^{PG}=1^{-+}$, G-even
(the quantum numbers are relative to those of the ground state
$I_{g.s.}^{PG}=0^{++}$ for $N/2$ even $I_{g.s.}^{PG}=0^{--}$ for $N/2$
odd).  The lattice operators
with the correct quantum numbers that create these states at zero momentum, 
acting on $|g.s.>$, and the corresponding states, read
\begin{eqnarray}
S&=&R+L\quad,\quad|S>=|0^{--}>=S|g.s.>\\
T_{+}&=&(T_{-})^{\dagger}=R^{(12)}+L^{(12)}
\label{ta}\\
T_{0}&=&\frac{1}{\sqrt{2}}(R^{(11)}+L^{(11)}-R^{(22)}-L^{(22)})
\label{t0}\\
|T_{\pm}>&=&|1^{-+},\pm1>=T_{\pm}|g.s.>
\label{tpm}\\
|T_{0}>&=&|1^{-+},0>=T_{0}|g.s.>
\label{to}
\end{eqnarray}
$R^{(ab)}$ and $L^{(ab)}$ in (\ref{ta},\ref{t0}) are the right and left
hopping operators ($L^{(ab)}=(R^{(ab)})^{\dagger}$)
$$
R^{(ab)}=\sum_{x=1}^{N}\psi_{a,x+1}^{\dagger}e^{iA}\psi_{b,x}
$$
Due to the mapping on the Heisenberg model
the norm of these states can be easily computed
\begin{eqnarray}
<S|S>&=&<g.s.|S^{\dagger}S|g.s.>=-4<g.s.|H_{J}|g.s.>\nonumber\\
<T_{+}|T_{+}>&=&\frac{2}{3}(N+<g.s.|H_{J}|g.s.>)\nonumber
\end{eqnarray}
$<T_{0}|T_{0}>=<T_{-}|T_{-}>=<T_{+}|T_{+}>$, 
$|g.s.>$ is normalized to the unity.

The isosinglet energy up to the second order is
$
E_{S}=E_{S}^{(0)}+\epsilon^{2}E_{S}^{(2)}
$,
where one has
\begin{eqnarray}
E_{S}^{(0)}&=&\frac{<S|H_{0}|S>}{<S|S>}=1\\
E_{S}^{(2)}&=&\frac{<S|H_{h}^{\dag}\Lambda_{S}H_{h}|S>}{<S|S>}
\label{secordex}
\end{eqnarray}
with
$
\Lambda_{S}=\frac{\Pi_{S}}{E_{S}^{(0)}-H_{0}}
$
and $1-\Pi_{S}$ a projection operator onto $|S>$.
Using the commutators
$
[H_{0},(\Pi_{S}H_{h})^n S]=(n+1)(\Pi_{S}H_{h})^n S
$,\
($n=0,1,\dots$), that hold when acting on $|g.s.>$, 
Eq.(\ref{secordex}) can be written in terms of spin correlators as
$$
E_{S}^{(2)}=E_{g.s.}^{(2)}+4-\frac{\sum_{x=1}^{N}
<g.s.|\vec{S}_{x}\cdot \vec{S}_{x+2}-\frac{1}{4}|g.s.>}{<g.s.|H_{J}|g.s.>}
$$

At the zeroth perturbative order the pseudoscalar triplet is completely 
degenerate with the isosinglet $E_{T}^{(0)}=E_{S}^{(0)}=1$. 
The second order energy of the states (\ref{tpm}) and 
(\ref{to}) can also be computed with an analogous procedure
\begin{eqnarray}
&&E_{T}^{(2)}=E_{g.s.}^{(2)}-\Delta_{DS}(T)-\nonumber\\ 
&&\frac{4<g.s.|H_{J}|g.s.>+5\sum_{x=1}^{N}
<g.s.|\vec{S}_{x}\cdot \vec{S}_{x+2}-\frac{1}{4}|g.s.>}{N+<g.s.|H_{J}|g.s.>}
\nonumber
\end{eqnarray}
where, introducing the operator
$
\vec{V}=\sum_{x=1}^{N}\vec{S}_{x}\wedge \vec{S}_{x+1}
$,
\begin{eqnarray}
\Delta_{DS}(T_{\pm})&=&12
\frac{<g.s.|(V_{1})^2|g.s.>+<g.s.|(V_{2})^2|g.s.>}
{N+<g.s.|H_{J}|g.s.>}\nonumber\\
\Delta_{DS}(T_{0})&=&12
\frac{2<g.s.|(V_{3})^2|g.s.>}{N+<g.s.|H_{J}|g.s.>}\ .\nonumber
\end{eqnarray}
The v.e.v. of each squared component of $\vec{V}$ on 
the rotationally invariant singlet $|g.s.>$ give the same contribution:
$\Delta_{DS}(T_{\pm})=\Delta_{DS}(T_{0})$, so that the 
triplet states (as in the continuum) have a degenerate mass gap. We verified 
this by direct computation on finite size systems. 
We used also finite size systems computations to demonstrate 
that $\Delta_{DS}$ is of zeroth order in N.
The excitation masses are given by $m_{S}=E_{S}-E_{g.s.}$ and 
$m_{T}=E_{T}-E_{g.s.}$. 
Consequently, the ($N$-dependent) ground state energy terms 
appearing in $E_{S}^{(2)}$ and $E_{T}^{(2)}$ cancel and what is left 
are only $N$ independent terms.
This is a good check of our computation, being the mass an intensive quantity.

The novel idea put forward in this letter is that in the strong
coupling, all the low-lying excitations of the Schwinger model can be
identified with those of the Heisenberg model and the mass spectrum of
the 2-flavor Schwinger model is computable in terms of spin-spin
correlators of the Heisenberg model. This shows that in strong coupling
there is an exact mapping between the gauge theory and the quantum
antiferromagnetic Heisenberg chain. 
Such a mapping was conjectured in~\cite{sem,hos}.
As evidenced in
\cite{korepin,coraff}, the explicit evaluation of the pertinent spin-spin
correlators is far from being trivial. It is an interesting problem
in its own right to derive expressions for spin correlators usable in
the evaluation of the mass spectrum of the 2-flavor Schwinger model.
We may devote our attention to this problem in a future publication.
This work is supported in part by the Natural Sciences and Engineering
Research Council of Canada, the Istituto Nazionale di Fisica Nucleare 
and M.U.R.S.T.

\end{document}